# Readership data and Research Impact

Ehsan Mohammadi[1], Mike Thelwall [2]

[1]School of Library and Information Science, University of South Carolina, Columbia, South Carolina, United States of America

[2]Statistical Cybermetrics Research Group, School of Mathematics and Computer Science, University of Wolverhampton, Wolverhampton, United Kingdom

**Abstract:** Reading academic publications is a key scholarly activity. Scholars accessing and recording academic publications online are producing new types of readership data. These include publisher, repository and academic social network download statistics as well as online reference manager records. This chapter discusses the use of download and reference manager data for research evaluation and library collection development. The focus is on the validity and application of readership data as an impact indicator for academic publications across different discipline. Mendeley is particularly promising in this regard, although all data sources are not subjected to rigorous quality control and can be manipulated.

**Subject terms**: readership; academic social network sites; reference managers; Mendeley

## 1.1 Introduction and Overview

The act of reading an academic publication is a key point at which knowledge is transferred from the author to someone else. With the prevalence of the web and social web, scholars now often read and register academic publications online, leaving electronic records of their activities. This readership data can reveal which outputs are used as well as giving insights how the scientific enterprise works. It is important to exploit such information to improve research evaluation practices and to investigate how science communication is evolving. The main advantage of using readership information rather than citations is that reading occurs before citing and can therefore give more timely information. A second advantage is that the reading public for academic research is wider than the citing public since it includes students, professionals and others.

        This chapter reviews research about download and readership data for academic outputs from the perspective of its value for research evaluation. It is mainly concerned with journal articles but books and conference papers are also briefly mentioned. The chapter has extended coverage of readership information from online reference managers, such as Mendeley, because

of their practical value for research assessment. It also discusses publisher usage data, since "using" an article in this context often means reading it.

The usage statistics component complements the *Usage Bibliometrics as a Tool to Measure Research Activity* chapter in this Handbook, which gives an extended case study of arXiv and analyses research evaluation at a more systemic level. The current chapter also discusses usage data from the academic social network sites ResearchGate and Academica.edu which are changing the way in which articles are discussed and shared. Previous works have reviewed usage bibliometrics (Kurtz & Bollen, 2010), readership metrics (Haustein, 2014), and social media metrics (Thelwall & Kousha, 2015a). This chapter updates the previous readership metrics chapter (Haustein, 2014) with newer topics and findings. It is intended for research evaluators, scientometricians and bibliometricians as well as those interested in recent changes in the scholarly communication ecosystem.

This chapter opens with a discussion of reading, terminology and research impact assessment to set the context for the ways in which readership data may be used. It continues with a discussion of online usage data from publishers and repositories, before analysing the effect of online reference managers on the availability of readership data and discussing social networks. Finally, the chapter discusses research evaluation applications.

## 1.2   Reading research: Background and terminology

Although knowledge has been traditionally communicated in human societies orally and by imitation, the written record is a cornerstone of modern science. Academics read scholarly publications to inform their current research, for current awareness purposes, to support their teaching, or to help them fulfil a professional advisory role related to their expertise. In addition, other professionals may read journals to inform their day-to-day practice. This is the case for medical doctors, who need to be aware of the latest developments in medical practice that are relevant to their role or expertise (Tenopir et al., 2007). For example, 73% of non-publishing Canadian physicians read journal articles (McAlister, Graham, Karr, & Laupacis, 1999), and hospital residents consider journal articles to be valuable sources of information for them (Schilling, Steiner, Lundahl, & Anderson, 2005). Information about which academic documents are read and by whom can help the librarians that buy them, the managers and policy makers that need to evaluate the impact of the research produced, and the scholars that investigate science itself.

Many different terms have been used for reading-related data, especially within electronic contexts. The most general is perhaps *usage*, which does not imply a reason why an item was accessed but is often employed in digital contexts as an umbrella term to describe accesses of all kinds of digital resource. It is helpful to use more specific terms, when justifiable, to aid interpretation.

The more explicit term *download* refers to a local copy being taken of an electronic resource, such as a journal article, whether by a human or a robot. The terms *hit* and *view* refer to online accesses of electronic information without necessarily downloading it. For example, a digital library visitor might *view* a webpage containing the title, abstract and metadata of an article and then *download* a full-text copy of the article to their local computer. The term *full-text download* (Haque & Ginsparg, 2009) can be used for emphasis. Although the term download usually connotes accessing a full text version of an article, some publishers provide non-downloadable full-text copies of articles and readers may view full-text articles online without downloading them to their local computer storage (e.g., Grabowsky, 2016).

The most specific term is *read/reader/readership*, which implies that the item accessed has been read by a human. This term can be justified for a data source if it is reasonable to believe that the items accessed will usually be read. For example, it would be reasonable to believe that a book is usually read when it is borrowed from a library. Thus, borrowing statistics could be claimed to be (non-exhaustive) reading indicators. In contrast, most accesses of documents on some websites may be from web crawlers and so it would not be reasonable to interpret download counts as readership counts (but see the paragraph below). Moreover, human users may also systematically download journal articles if they are concerned that their access will be interrupted before they know which articles to read (Emrani, Moradi-Salari, & Jamali, 2010). To describe download counts as a readership indicator, evidence would be needed to connect downloads to reading. Thus, if full text downloads from a specific source are likely to be of human origin and frequently lead to reading, then it would be reasonable to refer to full-text downloads as readership data.

More specifically, and following an argument for citations (van Raan, 1998), to be a readership *indicator*, download or access counts should positively correlate with reader numbers, even if not all downloads/accesses lead to reading. If this condition is fulfilled, then the counts convey information in an information theoretic sense about how often an item has been read. To give an example, a positive correlation implies that if article X has more accesses than Y then there is a greater than 50% chance that X has been more read than Y. In general, the higher the correlation, the higher this chance is. Aggregating sets of documents also, in general, increases this chance. Thus, a moderate positive correlation between accesses and reading would give a high change that research group A's outputs had been read more than research group B's outputs if they had been accessed more on average. In practice, the exact number of readers of any academic document is never known because of ways of accessing documents that cannot be tracked, such as from print copies of journals. Statistical correlations therefore need to be supported by arguments that the untracked readings are rare compared to the tracked readings or that the tracked readings are likely to be a relatively unbiased sample of all accesses. Alternatively, if the tracked accesses can be directly shown to correlate with research impact or value then this additional step is unnecessary.

## 1.3  *Readership data from libraries*

Two traditional sources of journal readership data are sales and library circulation information – either the number of libraries that hold a journal or circulation data for that journal. Both give journal-level rather than article-level evidence and are limited by the increasing share of reading that occurs online. For university libraries, the main journal readers are students, who are likely to be readers but not citers (Duy & Vaughan, 2006), and it is not clear whether they are also the main consumers of academic journals, or whether there are disciplinary differences in this. These students are likely to have different uses for academic journals and librarians need to consider this factor when analysing readership evidence to build collections (Rowlands & Nicholas, 2007).

Circulation information has been analysed for a long time (Gross & Gross, 1927) from practical (White & McCain, 1989) and mathematical modelling (Egghe & Rousseau, 2000) perspectives to support librarians (Kurtz & Bollen, 2010). Some examples illustrate the approaches used and give background for more current strategies.

The earliest usage records kept by libraries were lending or reshelving statistics and these have been used as proxies for usage or readership data for books and journals (King, Tenopir, & Clarke, 2006), even though they do not cover all ways in which they may be read (Davis, 2004). This data has been shown to correlate positively with Journal Impact Factors (JIFs), at least when comparing journals from the same discipline, and journals from specialisms that are similarly served by the library (Stankus & Rice, 1982; Tsay, 1998). This might be due to some readers subsequently citing articles in the journals or more cited journals tending to have more useful information. For example, *Nature* and *Science* target non-citing audiences in parallel with the acknowledged high quality of their scholarly content. In contrast, other journals primarily target a professional audience (e.g., nurses, librarians, lawyers) and may be less concerned to attract an academic readership. High correlations between online downloads and local readership information in one context give some evidence that downloads can be good indicators of readership (Emrani, Moradi-Salari, & Jamali, 2010).

Inter-library lending information and direct observations of library users can also give readership information, although the latter is too time consuming to be routinely used (Butkovich, 1996).

## *1.4* Research impact assessment

The publication of peer-reviewed research is a critical scholarly activity and the analysis of scholarly publications is important for assessing the research impact of scholars or teams in most disciplines. This assessment can be qualitative, quantitative or both. It can be combined with other sources of evidence or judgements and it can be made for formal evaluations, formative self-evaluations, or to investigate an aspect of science or science communication.

### 1.4.1 Peer review

Peer review is the central evaluation mechanism for modern science (Kostoff, 1997). The value of research is judged by experts from the same field because non-specialists are less able to understand the work or its contribution to scholarship. Ultimately, however, the scholarly community is collectively responsible to the governments or others that fund them and so experts may adjust their expectations in response. For example, this may lead a community to regard work as better if it offers societal benefits.

Peer review is also at the heart of the publication system, with journal articles being usually subject to critical evaluation before a decision is made about whether to accept them, and the same is often true for monographs (Wouters, 1997). It is also central to research funding applications (Jayasinghe, Marsh, & Bond, 2003), perhaps as one of a checklist of attributes to assess. In some countries, peer review is used to periodically assess research to allocate block research grants. In the UK, the Higher Education Funding Council for England (HEFCE) uses panels of field experts to evaluate the quality of research from government funded universities and other academic organizations to share out funding on a merit basis (HEFCE, 2011).

The main disadvantage of peer review is that it is slow and expensive, consuming a substantial amount of expert time (ABRC, 1990; Smith, 2006). It is also fallible because reviewers may be consciously or unconsciously biased against others' work based on personal characteristics, research competition or the research paradigm followed (Lee, Sugimoto, Zhang, & Cronin, 2013). Moreover, even expert reviewers can disagree about the merits of papers (Welch, 2014). Nevertheless, peer review is often the most credible single source of evaluation evidence for academic research.

Although peer review primarily assesses prior work rather than predicting future performance, experts can also assess applicants' plans if these are relevant for appointments and promotions.

### 1.4.2 Citation analysis

Citation-based indicators are the primary quantitative tools to evaluate research, whether on their own, to support human judgements or to cross-check reviewers' opinions (van Raan, 2000). Citation data may come from a recognized citation index, such as the Web of Science (WoS), Elsevier's Scopus, or Google Scholar. There are also specialist citation indexes for some fields and others with a national or linguistic scope, such as the Chinese Citation Index.

The best-known citation indicator is the Journal Impact Factor (JIF), which estimates the average number of citations to recently-published articles in a journal. The JIF is informally used as an indicator of the quality of a journal perhaps because of its simplicity and intuitive reasonableness. Nevertheless, there are numerous problems with the accuracy of the calculations and their ability to reflect impact (Egghe, 1988; Dellavalle, Schilling, Rodriguez, Van de Sompel, & Bollen, 2007). There are also major problems of over-interpretation, leading to inappropriate uses (Opthof, 1997; Boell & Wilson, 2010), such as those that ignore disciplinary differences.

The average number of citations per publication is known to vary by field and year and so it is not reasonable to compare the average citation count between groups of publications. Field normalized indicators, such as the Mean Normalized Citation Score (MNCS) (Waltman, van Eck, van Leeuwen, Visser, & van Raan, 2011ab) and Mean Normalized Log-transformed Citation Score (MNLCS) (Thelwall, 2017ab) solve this problem by normalizing citation counts for the publishing field and year so that a score of 1 always means citation impact equal to the world average.

For individual authors, the h-index has become popular (Hirsch, 2005), although it is biased towards senior researchers and male researchers.

All citation-based indicators suffer from several weaknesses. At a theoretical level, citations may reflect scholars acknowledging prior work that has influenced them (Merton, 1973), but they are not used systematically for this, can be influenced by spurious factors and can be negative (MacRoberts & MacRoberts, 1989; MacRoberts & MacRoberts, 1996). Moreover, citations only reflect knowledge advancement rather than wider contributions to academia or society (Kostoff, 1998; Merkx, Weijden Besselaar, & Spaapen, 2007). For example, academic publications can be used in education, the professions (Schloegl & Stock, 2004), and to inform health (Bennett, Casebeer, Kristofco, & Strasser, 2004; Lewison, 2002). The people affected in these cases may be thought of as "pure readers" in the sense of consuming academic outputs without subsequently citing them (Stock, 2009; Tenopir & King, 2000). Governments and research funders may explicitly state the need to consider non-academic impacts in their evaluations (NSF, 2013; Piwowar, 2013; Research Councils UK, 2011, para.1).

An important practical drawback of citation-based indicators is the considerable time that they take to accumulate. A research team might think of an idea, submit a grant proposal, get funded, carry out their research, submit a write-up to a journal, get it reviewed and accepted and then wait for the article to be published online or in a formal journal issue. This process might take several years and several more years would be needed before their article had attracted a reasonable number of citations from others who had read their work and then followed the same process to conduct related research. A citation window of three years is sometimes recommended for research evaluations (Abramo, Cicero, & D'Angelo, 2011). Thus, whilst academic evaluations often have the aim of predicting future research excellence so that it can be funded, or the promising academic can be appointed/promoted, in practice citation-based indicators reflect performance that occurred several years in the past.

## 2 Online access and download data

In response to the time delay and limited impact coverage problems of citation analysis as well as the expense of peer review, science policy makers and managers may seek alternative indicators to reflect wider types of research impact or to give earlier evidence of impact (Drooge, Besselaar, Elsen, & Haas, 2010; De Jong, Van Arensbergen, Daemen, Van der Meulen, & Van den Besselaar, 2011; Priem, Taraborelli, Groth, & Neylon, 2010).

Another important driver towards new indicators is the changing nature of scholarly communication, with increasingly diverse types of scholarly output being published online and valued even if they are rarely cited (Czerniewicz, Kell, Willmers, & King, 2014; Van der Vaart et al. 2013). In parallel, an increasing amount of scholarly communication takes place in public and online, leaving traces that may be used to evaluate its importance (Cronin, Snyder, Rosenbaum, Martinson, & Callahan, 1998).

Given that an academic publication must be accessed and read to be valuable but that not all readers are citers, it is logical to look to access and readership data for evidence of the wider impacts of publications (Haustein, 2014). This information can also ameliorate the time delay problem of citation analyses because an article must be read before any action can be taken based on its contents. Hence, unless the impact of an article is predicted from its metadata (e.g., from Journal Impact Factors), evidence of downloads or readership gives the earliest impact evidence. Away from research evaluation, this argument has also been made for library collection development. At the level of entire journals, readership and citation may be thought of as two separate, overlapping dimensions of the impact of research and giving librarians or evaluators information about both can help them to make more informed choices (Haustein, 2011; Haustein, 2012).

Although rarer in research evaluation contexts, the importance of readers is a natural concern for scholars, who may submit to journals partly based on the readership that they hope to gain (Tenopir & King, 2000), librarians who choose journals primarily to service their potential readers, and editors or publishers that monitor the overall audience or sales for a journal as an indicator of its value or health (Rousseau, 2002).

Readership data reflects something different from citations, even when only considering academic journal articles and restricting attention to academic readers. This is because there are some types of articles that are often read but rarely cited, such as a series of annual summaries of astrophysics (Kurtz, Eichhorn, Accomazzi, Grant, Demleitner, Murray, et al., 2005).

### 2.1.1 Online access and download data for journal usage assessment

Statistics about local online accesses of journals are the modern way for librarians to monitor readership levels, although they are not comprehensive. Libraries keep print copies of journals, researchers and faculty may subscribe to individual serials, and educators may photocopy articles for students. In addition, article preprints may be shared from institutional and subject repositories and home pages, as well as by email and post. These uses will not normally be recorded electronically by a library or publisher but online accesses nevertheless are likely to give more comprehensive and timely information than library circulation data. Online usage data may be able to separate out people that browse journals from those that read articles by accessing their full text, giving more substantial information than circulation data. Local data is more relevant to libraries than generic publisher data that covers all uses because each library serves a user community that has specific information needs – including research interests and educational provision. Moreover, local usage data seems to be more tied to reading since robots would presumably rarely access local library copies of articles, even though humans may still systematically download them (Emrani, Moradi-Salari, & Jamali, 2010).

Online usage information originates from the log file of a web server recording accesses of journal pages or downloads of electronic copies of articles. There are many technical pitfalls with this information, including accesses by robots and repeated accesses by individuals for spamming purposes or by accident, and so online accesses do not equate with human readership. Since publishers gain extra sales if their access statistics are higher, they do not have an incentive to clean their data from spurious downloads before delivering it to libraries or others. There have been initiatives to standardize the process of cleaning the data to ensure that compliant publishers generate credible and comparable final statistics for their end users. The main initiative for this is COUNTER (Counting Online Usage of NeTworked Electronic Resources), which standardizes the reporting of usage information (COUNTER, 2017; Pesch, 2015).

As discussed above, usage data is inaccurate because it is always an incomplete record of readership and there will also be unrecorded readers. In addition, there are other important limitations that apply to some or all contexts.

For all applications, the lack of contextual information with most usage data (Mendeley and questionnaire data are exceptions) is an important restriction. Librarians may consider usage by faculty to be more valuable to the university mission than uses by students, at least on an individual level, but can rarely distinguish between the two in their data. In addition, no current major data source gives evidence about how a document was used by the reader (Kurtz & Bollen, 2010). This is an advantage of citations, because, in theory, the reason for a citation can be deduced from the accompanying text.

### 2.1.2 Online access and download data for research evaluation

The shift to electronic publishing has led to the widespread availability of electronic access information at the level of individual articles, such as from publisher websites. This has made usage or readership data a practical source of evidence for research evaluations. In many cases usage and readership information can be used in a similar way to citations for impact assessment, although it has different advantages and limitations. It has also not become as generally accepted as citations for this purpose. For example, the Journal Impact Factor is much more widely reported and recognized than any of the proposed usage-based alternatives.

For research evaluations rather than collection development purposes, statistics that are available for entire journals but not individual articles are unhelpful, although monograph circulation data can help research evaluation in the arts, humanities and some social sciences (White, Boell, Yu, Davis, Wilson, & Cole, 2009).

Also for research evaluation purposes, the limited nature of local access data from individual libraries for journal articles can be resolved by forming a consortium of libraries to share data (Bollen & Van de Sompel, 2008) or by substituting publisher statistics. The former may be a realistic possibility for libraries that already have a common infrastructure to access electronic journals and so that data sharing can be added as an additional service rather than a completely new contribution.

Whatever the source of usage data, its principal advantage over citations for research evaluations is timeliness because usage logically comes before the publication of citations. A second advantage is scope because usage data includes, but does not differentiate, readers that do not cite the work. Hence usage data may provide a timelier source of impact evidence with a wider scope. The reason why it is rarely preferred to citations is that it is much easier to manipulate and so it is not credible enough for formal research evaluation purposes, even if from a COUNTER-compliant source. Nevertheless, it can be valuable for informal evaluations, self-evaluations and assessments of the science system as well as to cross-check the results of peer review or citation analysis.

Because of the accuracy limitations of usage data, it is important to assess whether it gives evidence of academic impact before it is used for article-level research evaluations. The primary strategy so far for this is to assess the extent to which article-level usage statistics correlate with citation counts. A perfect correlation cannot be expected because of the absence of pure readers from citation statistics, but a moderate or high correlation would suggest that the usage source assessed is not too affected by manipulation or fake data from robots. In contrast, a correlation close to zero would suggest that either there are many readers that have very different needs to citers or that the results have been intentionally or unintentionally manipulated.

Correlation analyses have mostly found moderate or high correlations between downloads and citations, which tends to confirm the value of usage data. A small study (n=153) of article downloads in their first week found a moderate correlation (Pearson r=0.5) with WoS citations five years later (Perneger, 2004). Similar correlations have been found for downloads of articles in the physics preprint server arXiv in the first six months and their citations after two years (Harnad & Carr 2006), for RePEc economics preprint downloads and citations (Chu, Krichel, & Blvd, 2007), for PLoS PDF downloads and citations (Yan & Gerstein, 2011. In contrast, a correlation of only 0.1 was found between early downloads (2 months) and later citations (25 months) for the fast, organic chemistry journal *Tetrahedron Letters*, suggesting that for this journal, early accesses represent a different type of use to citation (Moed, 2005). For downloads within Elsevier's ScienceDirect and Scopus citation counts, the two correlate in all disciplines at the level of journals and articles; early downloads also correlate with later citations. These correlations vary in strength by discipline, are lowest in the arts and humanities (0.2-0.3) and reach as high as 0.8 (life sciences). Despite this, the most downloaded articles tend to differ from the most cited articles for individual journals (Moed & Halevi, 2016). Confusingly, a study of Chinese journals found higher correlations between downloads and citations within the arts, humanities and social sciences than for other academic disciplines (Vaughan, Tang, & Yang, 2017).

At the level of journals, various download-based indicators have been defined in similar ways to the JIF, including the Usage Impact Factor (Bollen & Sompel, 2008) and the Download Immediacy Index (Wan, Hua, Rousseau, & Sun, 2010). Correlation tests have been used to help assess the value and validity of download-based indicators, with typically weaker results than at the level of individual articles (e.g., Wulff & Nixon, 2004), and with some negative correlations. Usage data for Rouen University Hospital digital library had a low positive correlation with JIFs in one study (Darmoni & Roussel, 2002), and correlations were not significantly different from zero for JIFs and Concordia University chemistry and biochemistry journal usage data (Duy & Vaughan, 2006). A comparison of JIFs with aggregate full text download data for a set of universities found low negative correlations, suggesting that journals most used by students (the main downloaders) were the least cited (Bollen, Van de Sompel, & Rodriguez, 2008; see also: Bollen and Sompel, 2008). Thus, whilst download data seems to reflect a high degree of scholarly impact at the level of individual articles, when articles are aggregated into journals, scholarly impact is substantially less important and download data may predominantly reflect educational value.

Electronic usage data can sometimes incorporate information about the origins of the users from the internet address of their computers. It is therefore possible to break down the readers of an article by institution and country and perhaps also organization type, if this data is made available by publishers or web server operators (Duin, King, & Van Den Besselaar, 2012; Kurtz, Eichhorn, Accomazzi, Grant, Demleitner, Murray, et al., 2005). This can reveal where articles and journals have had impact. This may be relevant to national funding bodies that want to demonstrate international impact or, conversely, want to make sure that the home nation rather than a competitor is benefiting from their investment.

### 2.1.3 Limitations of online usage data

As discussed above, usage data can include false hits, whether robot accesses or downloads by people that did not intend to read the article, and articles can be accessed from multiple print and online sources (Haustein & Siebenlist, 2011). These limitations apply unequally between journals and even between articles so it is not fair to compare the impact of articles using any source of download data. For example, one article's publisher download count may be half that of another because it is available free online from the author, or is available in print in addition to electronic versions (Anderson, Sack, Krauss, & O'Keefe, 2001). The main disadvantage of download counts from a research evaluation perspective is that they are easy to manipulate unless extensive measures are taken to protect them (Zimmermann, 2013). An additional disadvantage is that the data is not transparent because publishers do not share the identities of those that accessed an article and so authors and evaluators have no means of verifying downloads.

## 2.2 Readership data from online reference managers

In addition to manual methods to collect readership information, such as surveys, reader observation and reshelving information, and computerized methods, such as library, publisher or repository download statistics, the web has made possible an additional indirect method to find whether an article has many readers: online reference managers. A reference manager is a program that records metadata about some or all the resources that a person is interested in, typically to keep track of what they have read and to automatically generate reference lists for their documents, whether they are journal articles, conference papers, books or reports. Reference managers like EndNote, RefWorks, CiteULike, Connotea, Mendeley, Bibsonomy and Zotero all perform this role in different ways.

If it is assumed that reference manger users tend to record articles that they have read, then the collective databases of reference managers form a large source of information about what has been read by whom. Some reference managers do not share this information but others, such as Mendeley and Bibsonomy, do, and so reference manager records are an alternative source of readership information (Neylon & Wu, 2009).

At first glance, reference manager data is an unpromising source of readership evidence. Not all readers use reference managers and so they form an incomplete readership record. No reference manager is dominant and so if one is used as a data source then its information will be partial even with respect to all reference manager users. Reference manager users are likely to be a biased subset of all readers because technophobes might avoid learning a new program and people that do not write documents that include references would have little need for them.

Nevertheless, some reference managers have an advantage over download data: their information is freely available from a single source (rather than multiple publishers), they are not affected by multiple copies of articles being available (e.g., preprints in repositories) and they seem to give more definite evidence of readership than downloads because the latter could be from a crawler or a casual user. For this reason, they can be a more realistic data source for research evaluations than download data.

Data from reference managers that are also social websites and provide an Applications Programming Interface (API), such as Mendeley, CiteULike and Bibsonomy, falls within the scope of altmetrics (Holmberg, 2015). These are indicators derived from data harvested from social web sites via APIs. The altmetrics movement has led to the creation of many new indicators. Indicator companies, such as Altmetric.com, ImpactStory.org and Plum Analytics, systematically collect altmetric data (including from reference managers) and make it available to publishers, scholars and institutions (Wouters & Costas, 2012). Altmetric.com, for example, attempts to provide accurate and transparent article-level indicators (Liu & Adie, 2013). Although it includes readership data from Mendeley, it treats this as a secondary data source since it is not transparent (i.e., does not reveal the identity of readers). In addition, there are public initiatives to harvest and share altmetric data, such as one from PLoS (Fenner, 2013).

The promise of altmetrics is that it will deliver faster impact evidence that encapsulates wider types of impact (Priem, Piwowar, & Hemminger, 2012; Priem, Taraborelli, Groth, & Neylon, 2010). Within this, reference manager data fits as a fast and wider source of evidence since reference manager users may be students (e.g., Winslow, Skripsky, & Kelly, 2016) and other non-publishing article readers. Each altmetric has its own strengths and weaknesses, and potentially reflects a different type of impact. For example, tweet citations (Eysenbach, 2011) seem to reflect attention rather than impact and are probably the fastest indicator to accrue. Reference manager data can therefore be thought of as an alternative to download counts as a source of readership evidence, or as a type of altmetric to be analysed in parallel with other altmetrics.

### 2.2.1 Online reference managers: background

Online reference managers have broadly similar functions but each has its own software design, individual features and user demographics. The national, disciplinary and age composition of the adopters of each one is likely to be influenced by its age, national and disciplinary origins and the fit of its affordances within disciplinary missions. For example, most have Western origins and did not prioritise language support, which may have alienated potential users in China, Russia and Japan. User demographics are unknown for most, however. The descriptions below of some of the major social sites give a flavour of their differences but their capabilities evolve over time and so may have changed now. All are online social reference managers in the sense that they manage references, are online, and allow users to create a public profile.

- Bibsonomy (www.bibsonomy.org) manages web bookmarks as well as references and incorporates social features (Benz, Hotho, Jäschke, Krause, Mitzlaff, Schmitz, & Stumme, 2010; Jäschke et al., 2007; Ortega, 2016). Users typically have about 20% more references than bookmarks (Ortega, 2016). Probably like the other sites, most references are for journal articles (Borrego & Fry, 2012). Bibsonomy provides a copy of all its data free for researchers (www.kde.cs.uni-kassel.de/bibsonomy/dumps/).
- CiteULike (citeulike.org) is free, has the basic reference manager capabilities described above and allows users to annotate references and share them with others (Reher & Haustein, 2010). It also has discussion fora and blogs (Bogers & Van Den Bosch, 2008). Because of its communication and sharing capabilities it is also a type of academic social web site.
- Mendeley (Mendeley.com) is a free reference manager (Henning & Reichelt, 2008) that has been bought by Elsevier and offers social networking features, such as the ability to follow other users, as well as discussion groups and the ability for users to list their own publications. It therefore serves as an academic social network site as well as a reference manager, although its social features do not seem to be extensively used (Jeng, He, & Jiang, 2015). Mendeley offers a free and comprehensive API to access its readership data for all articles in its catalogue so that anyone can find out how many Mendeley users have recorded any given document within their libraries. Although Mendeley does not report which users have registered each document, it gives a breakdown of user types by status (e.g., undergraduate, professor, other professional), geographic location (country) and main academic discipline (Gunn, 2013).
- Zotero (www.zotero.org) is a free, open source reference manager that originated as a Firefox web browser plugin but is now available as a separate program. It has features to support group discussions and group reference sharing.

In addition to the above, RefWorks is a reference manager owned by ProQuest, and EndNote is owned by Thomson Reuters. Neither share readership data publicly at the time of writing.

### 2.2.2 Online reference managers: Coverage

Readership data from online social reference managers needs to be publicly available, or at least shared with researchers, and to have many records to be useful. If a site has few users, then these are likely to be a very biased subset of readers so the results may be misleading. For example, article readers tend to be from the same country as the authors (Thelwall & Maflahi, 2015) so any national readership biases will translate into international readership indicator biases. If most articles do not have a record of readers in the site, then its data is unlikely to be powerful enough for research evaluation purposes unless used on a large scale to compare average reader counts (e.g., using the EMNPC: Thelwall, 2017bd). Of the online reference managers sharing data, Mendeley has the widest coverage and probably the most users. It had records for 80% of PLoS articles compared to 31% for CiteULike (Priem, Piwowar and Hemminger, 2012) and indexed more *Nature* and *Science* articles (Li, Thelwall, & Giustini, 2012).

- Bibsonomy: Bibsonomy has much lower coverage of physics journal articles 2004-2008 than CiteULike and probably less than 1% (Haustein & Siebenlist, 2011). Journal articles comprise half (47%) of the items recorded, with conference papers (25%) and books (12%) also being common (Borrego & Fry, 2012).
- CiteULike: Most (65%) PloS Biology articles have a record in CiteULike (Fenner, 2013). Less than 3% of physics articles 2004-2008 are in CiteULike (Haustein & Siebenlist, 2011).
- Mendeley: Virtually all (97%) articles from *Journal of the American Society for Information Science and Technology* 2001-2011 (Bar-Ilan, 2012) and *PloS Biology* (95%) have a record in Mendeley (Fenner, 2013). Most (66%) PubMed articles 2010-2012 that are also in the Web of Science have a Mendeley record (Haustein, Larivière, Thelwall, Amyot, & Peters, 2014). For Scopus medical fields, 78% of articles had at least one reader (Thelwall & Wilson, 2016). Another study found high coverage for WoS Medicine 2008 articles (72%) but lower (about a third) for Physics, Chemistry, Engineering and Technology (Mohammadi, Thelwall, Haustein, et al., 2015). Less than half of recent social sciences WoS articles (44%) are in Mendeley, varying from Psychology (54%) to Linguistics (34%) and only 13% of humanities articles were indexed, from Education (34%) to Literature (4%) (Mohammadi & Thelwall, 2014). Nevertheless, 61% of Swedish humanities journal articles from 2012 were in Mendeley (Hammarfelt, 2014). Compared to other altmetrics from Altmetric.com, Mendeley had the highest coverage (63%) of a large sample of Web of Science articles (Zahedi, Costas, & Wouters, 2014). Very few books have records: only 7% of recent WoS science and medicine volumes (Kousha & Thelwall, 2016). Thus, whilst Mendeley has wide coverage overall and particularly for medicine, it is weak in the humanities and very weak in some disciplines and for books. This may have changed since the studies described here however.
- Zotero. No coverage information is available.

### 2.2.3 Online reference managers: Correlation with citation counts

When a new indicator is proposed for an aspect of research evaluation then the logical first method to assess whether it has any value is to calculate its correlation with citation counts on the basis that a positive result would be evidence that the data was not random and related to scholarly impact in some way (Sud & Thelwall, 2014). Even though a negative or zero correlation is also consistent with a new indicator reflecting a completely different type of impact, in practice most types of impact relate to each other to some extent and so this test is reasonable. There is extensive evidence of this type for Mendeley and a little for CiteULike. For Mendeley, readership counts correlate positively and moderately strongly with citation counts (and peer review judgements) in most fields, with the arts being the main exception.

- CiteULike records and citations have a significant positive correlation for *Science* and *Nature* (Li, Thelwall, & Giustini, 2012). Usage data dominated by CiteULike have low Spearman correlations (0.1) with Journal Impact Factors for physics journals (Haustein & Siebenlist, 2011)
- Mendeley records and citations have a significant positive correlation for *Science* and *Nature* (Li, Thelwall, & Giustini, 2012), for *PLoS ONE*, *PLoS Biology* and *PLoS Pathogens* articles (Priem, Piwowar, & Hemminger, 2012) and for selected genetics and genomics articles (Li & Thelwall, 2012). Mendeley readers have a moderate overall correlation (0.5) with WoS article citations (Zahedi, Costas, & Wouters, 2014). For PubMed articles 2010-2012 in WoS, Spearman correlations between Mendeley readers WoS citations were positive and statistically significant in all broad disciplines except the Arts. They varied from 0.2 (Humanities) to 0.6 (Engineering and Technology), with an average of 0.5 (Haustein, Larivière, Thelwall, Amyot, & Peters, 2014). For WoS articles from 2008, five social science fields had Spearman correlations of 0.4-0.6 and five humanities fields had Spearman correlations of 0.4 or 0.5 (Mohammadi & Thelwall, 2014; see also: Thelwall, in press). Similar correlations were found for science and medicine fields (0.4 or 0.5) except for Engineering and Technology (0.3) (Mohammadi, Thelwall, Haustein, & Larivière, 2015). Within Medicine the correlations later (and for narrower fields) rose to 0.7 (Thelwall & Wilson, 2016). The most systematic analysis so far checked 325 narrow Scopus fields, finding strong positive correlations in almost all (Thelwall, 2017f). For books, correlations between Mendeley reader counts and citations are about 0.1 (Kousha & Thelwall, 2016). Engineering conference papers have a very low correlation with citation counts (Aduku, Thelwall, & Kousha, 2017).

A more direct source of evidence of the value of readership counts is their correlation with peer review scores. Peer review judgements are impractical to obtain for large samples of articles unless the data is a by-product of a research evaluation. For articles published in 2008 and submitted for evaluation by subject experts in the UK's Research Excellence Framework (REF) 2014, correlations between Mendeley reader counts and expert ratings in 33 of the 36 fields examined were positive, with the highest being for clinical medicine (0.4) and the lowest for Music, Drama, Dance and Performing Arts (-0.1) (HEFCE, 2015). Given that these articles were selected by their authors for being high quality, the correlations are likely to substantially underestimate the underlying level of agreement between peer judgement and Mendeley reader counts and so this is strong evidence that in most fields Mendeley reader counts reflect the quality of journal articles. A weaker corroborating piece of evidence is that UK clinical guideline references have more Mendeley readers than do comparable articles (Thelwall & Maflahi, 2016)

### 2.2.4 Online reference managers and reading

References can be logged within a reference manager by users that have not read them (Weller & Peters, 2012) or as part of training exercises (MacMillan, 2012) and so it is not clear that it is reasonable to characterize reference manager data as "readership counts". The best way to find out why users enter reference data is to ask them. A survey of Mendeley users found that most (85%) added articles to cite them, but many also added articles for professional (50%) or teaching (25%) purposes. Moreover, most added articles that they had read or intended to read. Thus, Mendeley readership data clearly represents readership and a wider type of impact than scholarly impact, although mainly still within a broad academic context (Mohammadi, Thelwall, & Kousha, 2015). Some articles are added for educational reasons, having many readers but few citations (Thelwall, 2017c).

Since undergraduates use reference managers, it is logical to look to readership data for evidence of educational reading. This is undermined by evidence that undergraduates and researchers tend to register similar articles (Thelwall, 2017a)

### 2.2.5 Online reference managers: Reader types and demographics

Readers of research can originate from any country in the world, from any discipline, from any academic status. They can also be professionals using the information for their work or could be members of the public with an interest in a specific topic or fact. Within these groups, some read more academic research than others, and even when considering academic researchers alone, younger researchers read and cite more (Larivière, Sugimoto, & Bergeron, 2013; Tenopir, King, Spencer, & Wu, 2009). Undergraduates sometimes read scientific papers but their reading is often directed by their lecturers (Korobili, Malliari, & Zapounidou, 2011; Whitmire, 2002). To interpret the type of impact reflected by readership data, it is therefore important to investigate the nature of people that use online reference managers. Partial information is known for Bibsonomy and Mendeley.

In terms of geography, almost half of all Bibsonomy users are from Germany (Ortega, 2016), undermining its value for general impact assessment. Probably all the major services have relatively low numbers of users from China and from countries with little scientific publishing or a low level of technology use in universities.

In terms of work status, Mendeley includes substantial numbers of undergraduates and Master's students and few non-academic users. In science, it is dominated by young users: PhD students, postgraduates and postdoctoral researchers (Mohammadi, Thelwall, Haustein, & Larivière, 2015). In contrast, successful senior researchers seem to avoid it (Mas-Bleda, Thelwall, Kousha, & Aguillo, 2014) and so there is an age/seniority bias.

### 2.2.6 Online reference managers: Timeliness

Mendeley readers appear about a year before citations, on average. For four library and information science (LIS) journals, the number of citations reaches the number of readers after about seven years (Maflahi & Thelwall, 2016). A similar pattern of initially higher readership counts than citation counts has been found for 50 fields, although the number of years needed for citations to overtake readers varies by discipline (Thelwall & Sud, 2016). Early Mendeley readership counts are also better predictors of later high citation counts than are journal impact factors or citations (Zahedi, Costas & Wouters, 2015). All this evidence supports the conclusion that Mendeley readership counts give statistically stronger impact evidence than citation counts in the first few years after publication.

It is common for articles to have Mendeley readers as soon as they are formally published because of the prior sharing of preprints (Maflahi & Thelwall, 2018). This makes it possible to conduct evaluations of them immediately upon publication, if these evaluations do not require the statistical power of high average readership counts.

Most importantly, the higher number of Mendeley readers than citations in the year following publication makes Mendeley reader counts correlate more strongly than citation counts with peer review judgements of the quality of journal articles (HEFCE, 2015).

### 2.2.7 Online reference managers: Research evaluation applications

Readership data is useful for research evaluation applications where timeliness is important and there is little risk of deliberate manipulation of the data. This excludes formal exercises where those evaluated are told the data sources in advance but allows their use for more routine academic progress monitoring.

Mendeley readership counts are useful for national-level evaluations for governments to compare their progress against that of their competitors. The goal of such evaluations is to inform policy decisions or to assess the effect of recent policy changes. For the latter case in particular, timely data is essential. Mendeley readership data is preferable to citations because of its timeliness but has the limitation that it is influenced by different levels of national uptake from its users. This is a problem because of the tendency for people to read articles from their own country. It is possible to circumvent this issue with a modelling approach to measure the amount of bias in caused by the readership demographics and then correct for them, but this strategy is imperfect because it requires assumptions or information about the evolution of national uptake of the site over time (Fairclough & Thelwall, 2015).

Funding councils are also logical users of readership data. These may monitor the average impact of the research that they fund to identify which funding streams are most successful and whether the average impact of their funded research has changed over time. Web indicators can be useful for these because the time lag of citations would delay decisions about changing ineffective funding strategies (Thelwall, Kousha, Dinsmore, & Dolby, 2016) as well as for evidence of societal research impact (Dinsmore, Allen, & Dolby, 2014). National biases in uptake and the bias towards younger users are relatively minor problems for this and so Mendeley readership data is a better source than citations (e.g., Thelwall, 2017b), although it does not reflect societal benefits or many professional uses.

One recent application harnesses readership data purely for its early impact evidence in comparison to citation counts, emphasizing the importance of the publication lag for citations. It used reader counts for early evidence of the success of an article promotion strategy in a randomised controlled trial (Kudlow, Cockerill, Toccalino, Dziadyk, Rutledge, et al., 2017).

### 2.2.8 Illustration of Mendeley data

Three documents were compared with Mendeley data to illustrate some of the features of Mendeley and some of the factors that should be considered when interpreting its data. Three important scientometric papers were selected for this comparison. The first is an old Nature article discussing citation analysis from the perspective of non-scientometricians concerned about its uses. The second is the well-known Leiden Manifesto. The third is an article about altmetrics (Table 1).

All three articles have high numbers of readers and Google Scholar citations. Other factors being equal, older articles should be more cited so it seems likely that the second and third articles, from 2015, will eventually be more cited than the first one, from 2002. The two newer articles already have more Mendeley readers than the first article (Counting house). This is partly because Mendeley identified readers before citations and so newer articles take less time to catch up with older articles in terms of reader counts. It is also partly because the Counting house article was published years before Mendeley was released so its peak years of use would have preceded the existence of substantial numbers of Mendeley users.

Table 1. Mendeley readers and Google Scholar citations for three scientometrics articles.

| Title and authors | Year | Readers | GS cites | Reads/Cites |
|---|---|---|---|---|
| Citation analysis: The counting house *by* Adam, D. | 2002 | 173 | 576 | 0.30 |
| The Leiden Manifesto for research metrics *by* Hicks, D., Wouters, P., Waltman, L., De Rijcke, S., Rafols, I. | 2015 | 634 | 474 | 1.34 |
| Do altmetrics correlate with citations: Extensive comparison of altmetric indicators with citations from a multidisciplinary perspective *by* Costas, R., Zahedi, Z., Wouters, P. | 2015 | 389 | 251 | 1.55 |

Mendeley data includes users' professions (Figure 1). Most strikingly, the altmetric article is most used by librarians. Presumably this is due to the proliferation of altmetrics in publisher websites. In contrast, professors seem to be more concerned with traditional citation-based indicators.

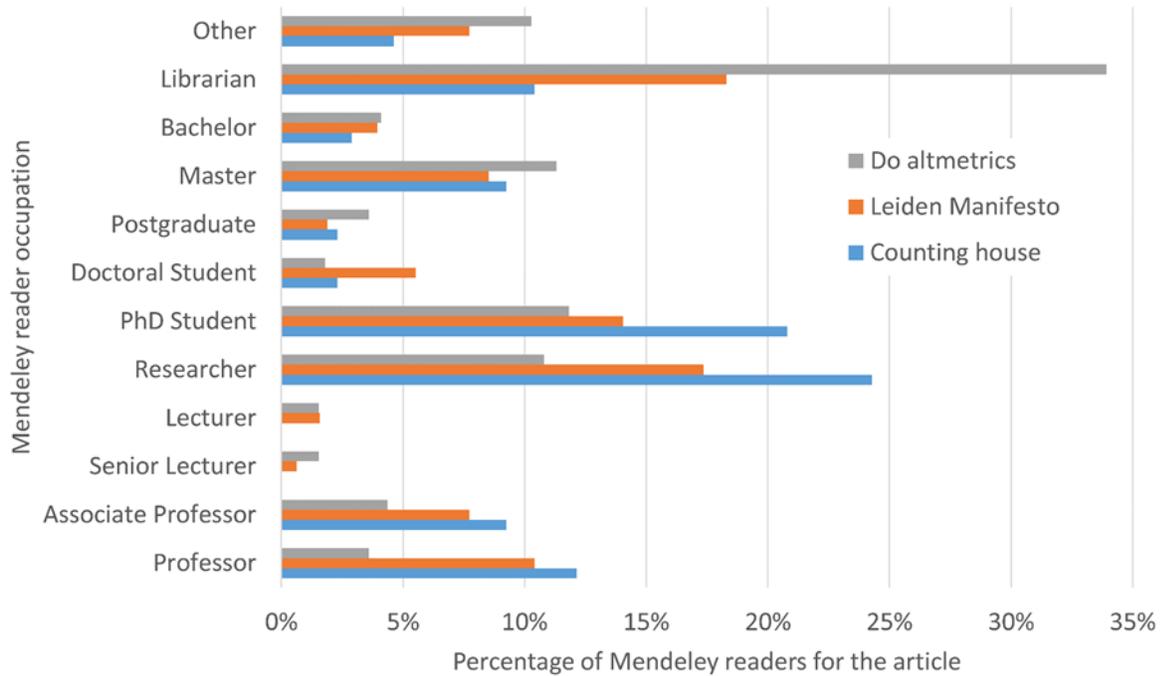

Figure 1. All Mendeley reader occupations for three scientometric articles.

Mendeley data includes users' declared country of origin or work (Figure 2). It seems that some countries that are taking citation analysis seriously, such as Brazil, are not concerned with altmetrics. In contrast, Canada and Netherlands seem to be more interested in altmetrics than citation analysis although both countries have active researchers working in both areas. The Counting house article seems to be particularly influential in the USA but it is not clear why.

Figure 2. Countries of Mendeley readers for countries with at least 10 readers.

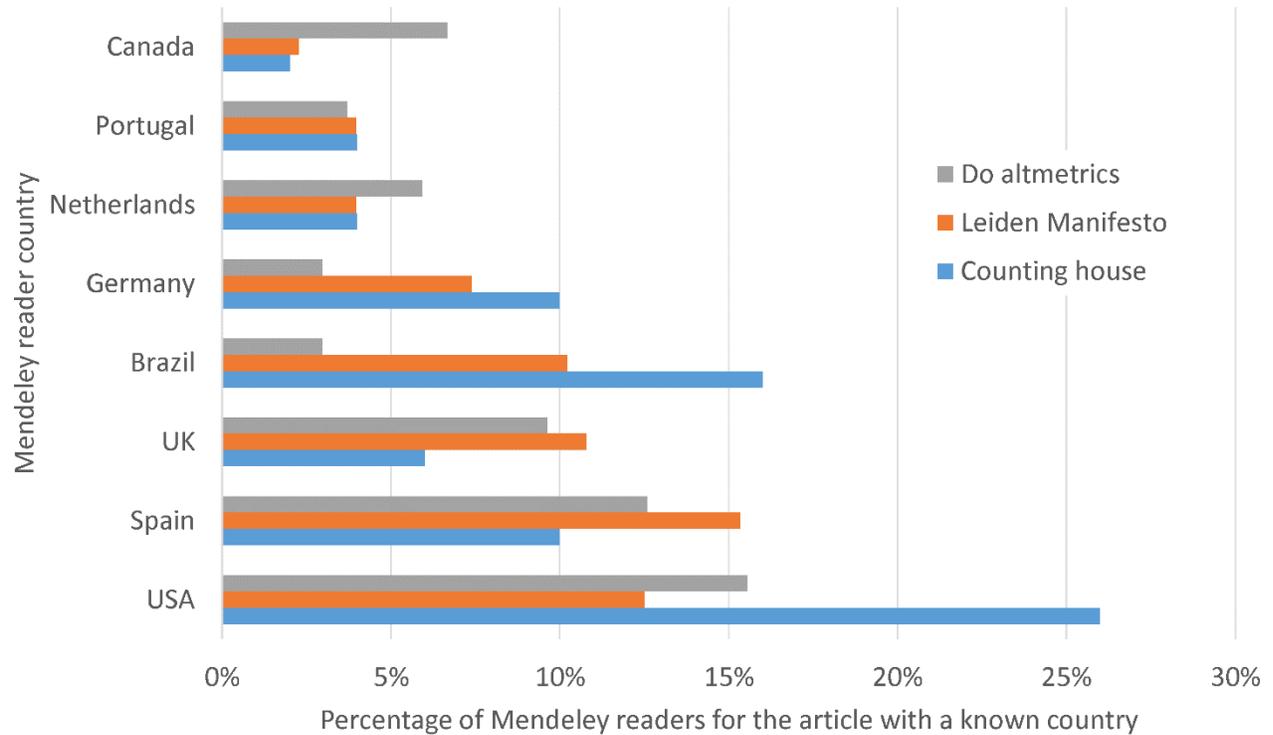

There are substantial disciplinary differences in the uptake of the articles (Figure 3). The altmetrics article has attracted the most attention in the Social Sciences and Computer Science, although both categories might be due to library and information science researchers since this field falls within both. The citation analysis articles are of interest in the Agricultural and Biological Sciences. This is unsurprising given the origins of the San Francisco Declaration on Research Assessment (DORA) within the life sciences (the American Society for Cell Biology), indicating an unease with misuses of citation analysis within this discipline. Figure 3 also confirms that all three articles have attracted substantial interest outside of their home disciplines.

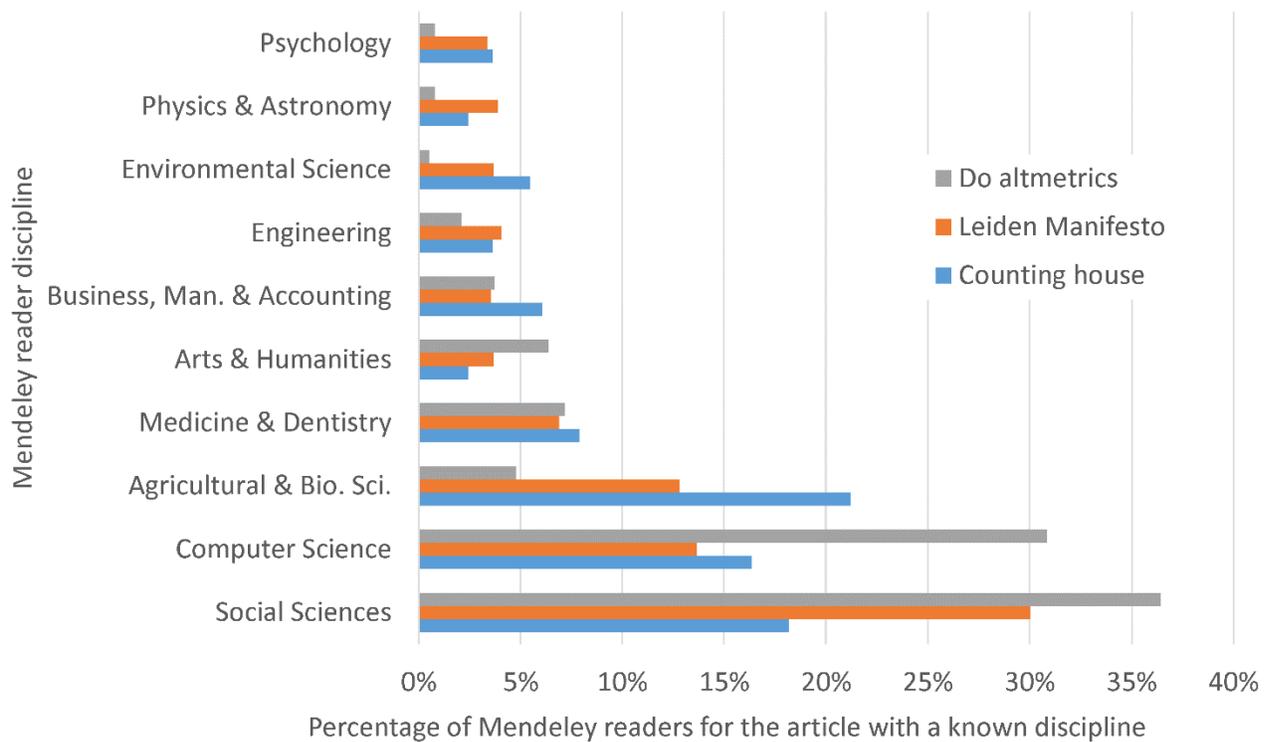

Figure 3. Disciplines of Mendeley readers for the top 10 disciplines.

### 2.2.9 Investigating science with online reference manager data

An interesting application of readership data is to track the flow of knowledge between fields. There is a long tradition of using citations to track knowledge flows by interpreting a citation from an article in field A to a paper in field B as knowledge flowing from B to A (Small, 1997). The same is possible for readership data when the domains of the readers of an article are known. The advantages of using readership data for this are timeliness and its ability to capture slightly wider impact because of the inclusion of students. A lot of data is needed to give good results however, which was a problem with one CiteULike study (Jiang, He, & Ni, 2011). An investigation comparing knowledge flows based on Mendeley readership data with citation-based knowledge flows found differences suggesting that researchers in some fields, including Business, read widely but cited narrowly (Mohammadi & Thelwall, 2014).

A related application is the discovery of research clusters by identifying groups of articles read by the same user and then clustering them based on co-readership information (Kraker, Körner, Jack, & Granitzer, 2012), although this seems to be no longer possible with Mendeley.

Readership data has also been used to investigate academics through their reference lists, when these are public (Borrego & Fry, 2012), to evaluate journals through the extent to which they are read (Haustein & Siebenlist, 2011), and to support literature search systems (Heck, Peters, & Stock, 2011).

### 2.2.10 Advantages and disadvantages of reference manager data compared to citation counts

The timeliness, wider impact, and reader demographic information advantages of readership data from all sources have already been mentioned, as have the disadvantages that it is sometimes not transparent and always open to manipulation, with a biased user base. Some additional factors are important to consider when evaluating readership data. The advantages are listed first.

- Traditional citation indexes, such as WoS and Scopus, have national biases and limits in coverage (Archambault, Campbell, Gingras, & Larivière, 2009) whereas there are no restrictions on the articles that may be added to reference managers.
- Readership data is often free whereas citation indexes, except Google Scholar and Microsoft Academic, tend to charge for access.
- Readership data is relatively easy to access on a large scale from sites with an API. For example, the free Webometric Analyst software can download Mendeley records via its API from the articles DOIs and/or metadata.
- Readership data tends to be more numerous than citation counts (e.g., Thelwall, 2017e), except for older articles, and tests using it can therefore be statistically more powerful.

There are also additional disadvantages with readership data.

- Whereas, in theory, it is possible to find out how a work has been cited by reading the text accompanying the citation, references are rarely annotated with information that reveals why a publication was selected. Reviews are annotated readings and these are available from sites like Goodreads and Amazon for books (Kousha & Thelwall, 2016; Kousha, Thelwall, & Abdoli, 2017).
- Despite the recognised national biases in citation indexes, Mendeley readership data seems to be more nationally biased than citation counts (Mas-Bleda & Thelwall, 2016).
- Some altmetric sources of readership data can give inconsistent results (Fenner, 2014) and there is a need for standardisation between data providers and sources (Konkiel, 2013).
- Younger readers are more represented in Mendeley (Mohammadi, Thelwall, Haustein, & Larivière, 2015) and the share of younger readers may vary by narrow field and publication year.
- Differences in adoption levels and behaviours across disciplines is a complicating factor when interpreting the results of any multidisciplinary analysis (Wouters & Costas, 2012).
- Some publication information entered by users to record their references is incomplete, leading to missed data (Bornmann & Haunschild, 2015). This may be more frequent for documents with mathematical titles or in languages that are not represented by the ASCII character set.

## 2.3  *Usage data from academic social network sites*

The online environment for science communication is continually evolving and usage data is now not only available from publishers, academic repositories and reference managers but also from some academic social network sites. Both Academia.edu and ResearchGate allow members to log their own papers and add them to their profile pages (as Mendeley also does now). They also provide usage data on these records in the form of download or view counts. They differ from reference managers by focusing on each author's own publications rather than their references of (presumably) mainly other scholars' works. Thus, their usage data has essentially the same nature as the download and access statistics of publishers or repositories, even though their appearance is more like Mendeley. Academic social network sites are in competition with publishers as sources of published academic research and, because of this, undermine the comprehensiveness of publisher data, apparently irrespective of copyright concerns (Jamali, 2017).

For research evaluation purposes, academic social network sites are not good sources of usage indicators because they have an incomplete collection of articles and do not make their usage data easily available for researchers. Nevertheless, they are important because they have many mamebers and their scores are apparently taken seriously by many researchers (Jordan, 2015, 2017; Orduna-Malea, Martín-Martín, Thelwall, & López-Cózar, 2017).

ResearchGate article views correlate positively with Scopus citations and seem to reflect a wider set of users than publishing academics, putting them on a par with other sources of usage data (Thelwall & Kousha, 2017). ResearchGate also provides citation counts for uploaded articles by extracting citations from all articles uploaded to the site. Although it indexes less citations than Google Scholar, it finds more early citations than WoS and Scopus, suggesting that many authors upload preprints to the site (Thelwall & Kousha, 2017b). There are differing national levels of uptake of ResearchGate, which will bias its data, but despite being a type of web social network site, its data does not seem to favour younger users (Thelwall & Kousha, 2015b).

There is less research about Academia.edu, but, like ResearchGate, its scores seem to favour senior academics. They also tend to favour women, perhaps due to their greater communication expertise in the social web (Thelwall & Kousha, 2014).

## 2.4 Summary

This chapter has summarized research into readership data, including usage data, with a focus on research evaluation applications but also covering collection development applications. In theory, this data is preferable to citation counts because it captures more uses of scholarly documents, such as from students and professionals. Although there is a little evidence to support his conjecture, readership data seems to primarily reflect scholarly uses in most fields. Both readership data from reference managers and usage (download/view) data from publishers have the advantage of giving early impact evidence compared to citations because of the delays associated with the publication cycle. This is due to articles being read a year or more before the citations generated by the reading, if any, appear in a citation index. Nevertheless, both download data and reference manager data can be manipulated and whilst they are useful for informal evaluations and investigations into science itself, they should not be used for formal evaluation when those assessed can influence the data.

Readership data from reference managers has the additional promise that it can reveal something about the demographics of the readers, including their discipline, nation and job type. This can help with investigations of science communication. Download data in many cases has the practical limitation that a set of articles may originate from many different publishers, which complicates accessing it and the data may not be fully comparable. In contrast, it is reasonable to collect reference manager readership data from a single site, such as Mendeley via its API, making it a practical source of readership information.